\pdfoutput=1

\documentclass{jfm}
\usepackage{amsmath,gensymb}
\usepackage{color}
\usepackage{graphicx}
\usepackage{natbib}
\graphicspath{{figs/}}

\ifCUPmtlplainloaded \else
  \checkfont{eurm10}
  \iffontfound
    \IfFileExists{upmath.sty}
      {\typeout{^^JFound AMS Euler Roman fonts on the system,
                   using the 'upmath' package.^^J}%
       \usepackage{upmath}}
      {\typeout{^^JFound AMS Euler Roman fonts on the system, but you
                   dont seem to have the}%
       \typeout{'upmath' package installed. JFM.cls can take advantage
                 of these fonts,^^Jif you use 'upmath' package.^^J}%
      }
  \else
  \fi
\fi

\ifCUPmtlplainloaded \else
  \checkfont{msam10}
  \iffontfound
    \IfFileExists{amssymb.sty}
      {\typeout{^^JFound AMS Symbol fonts on the system, using the
                'amssymb' package.^^J}%
       \usepackage{amssymb}%

      }{}
  \fi
\fi

\ifCUPmtlplainloaded \else
  \IfFileExists{amsbsy.sty}
    {\typeout{^^JFound the 'amsbsy' package on the system, using it.^^J}%
     \usepackage{amsbsy}}
    {}
\fi

\newcommand\cm{\nobreak\mbox{$\;$cm}}

\newcommand\cmcubepersec{\nobreak\mbox{$\;$cm$^3$\,s$^{-1}$}}
\newcommand\cmsqpersec{\nobreak\mbox{$\;$cm$^2$\,s$^{-1}$}}
\newcommand\cmpersecsq{\nobreak\mbox{$\;$cm\,s$^{-2}$}}
\newcommand\cmpersec{\nobreak\mbox{$\;$cm\,s$^{-1}$}}

\newsavebox{\astrutbox}
\sbox{\astrutbox}{\rule[-5pt]{0pt}{20pt}}

\newcommand{\mathnotation}[2]{\newcommand{#1}{\ensuremath{#2}}}

\renewcommand{\l}{\left}
\renewcommand{\r}{\right}
\mathnotation{\pd}{\partial}      % partial derivative

\mathnotation{\uc}{u}             % radial velocity component
\mathnotation{\wc}{w}             % vertical velocity component
\mathnotation{\rc}{r}             % radial coord
\mathnotation{\zc}{z}             % vertical coord
\mathnotation{\grav}{g}           % gravitational acceleration
\mathnotation{\gravt}{(g\cos\ang)}% vertical gravitational acceleration
\mathnotation{\visc}{\nu}         % viscosity

\mathnotation{\ang}{\theta}       % inclination angle

\mathnotation{\ua}{v}             % vertically-averaged horizontal velocity
\mathnotation{\h}{h}              % thickness
\mathnotation{\G}{G}              % d'less gravity
\mathnotation{\Q}{Q}              % flowrate
\mathnotation{\q}{q}              % flowrate/2pi
\mathnotation{\Uc}{U}             % jet velocity
\mathnotation{\rr}{r}             % radial coord
\mathnotation{\ff}{f}             % vertical profile
\mathnotation{\ci}{c_1}           % vertical profile constant c1
\mathnotation{\cii}{c_2}          % vertical profile constant c2

\mathnotation{\rjump}{R_{\text{jump}}}
\mathnotation{\rsing}{R_{\text{sing}}}
\mathnotation{\rcjump}{\rc_{\text{jump}}}
\mathnotation{\rcsing}{\rc_{\text{sing}}}

\title[Hydraulic jumps on an incline]{Hydraulic jumps on an incline}

\author[J.-L. Thiffeault and A. Belmonte]%
{
  J\ls E\ls A\ls N\ls -- L\ls U\ls C\ns
  T\ls H\ls I\ls F\ls F\ls E\ls A\ls U\ls L\ls T$^{1,2}$
\and
  A\ls N\ls D\ls R\ls E\ls W\ns
  B\ls E\ls L\ls M\ls O\ls N\ls T\ls E$^3$
}

\affiliation{
  $^1$ Department of Mathematics, University of Wisconsin, Madison,
  WI 53706, USA\\[\affilskip]
  $^2$ Institute for Mathematics and its Applications, University of
  Minnesota, Minneapolis, MN 55455, USA\\[\affilskip]
  $^3$ The W. G. Pritchard Laboratories, Department of Mathematics, Penn
  State University, University Park, PA 16802, USA}

\pubyear{2010}
\volume{XX}
\pagerange{xx--xx}
\date{}

\begin{document}

\maketitle

\begin{abstract}
  When a fluid jet strikes an inclined solid surface at normal
  incidence, gravity creates a flow pattern with a thick outer rim
  resembling a parabola and reminiscent of a hydraulic jump. There
  appears to be little theory or experiments describing simple aspects
  of this phenomenon, such as the maximum rise height of the fluid
  above the impact point, and its dependence on jet velocity and
  inclination angle. We address this with experiments, and present a
  simple theory based on horizontal hydraulic jumps which accounts for
  the rise height and its scaling, though without describing the shape
  of the parabolic envelope.
\end{abstract}


\section{Introduction}

The description of the hydraulic jump arising from a jet striking a
solid surface has been a rich source of fluid dynamical problems for
decades.  The first fairly complete description is usually attributed
to~\cite{Watson1964}, and his theory has been refined and improved by
many authors (for
example~\cite{Bohr1993,Godwin1993,Bohr1997,Brechet1999,Chang2001,Bush2003}).
Experimentally, there are several variants of the problem, such as
where the jet strikes a horizontal plate at an oblique
angle~\citep{Sparrow1980,Rubel1981,Kate2007}, or where the plate
is moving~\citep{Gradeck2006,Kate2009}.

\begin{figure}
  \centerline{\includegraphics[width=.6\textwidth]{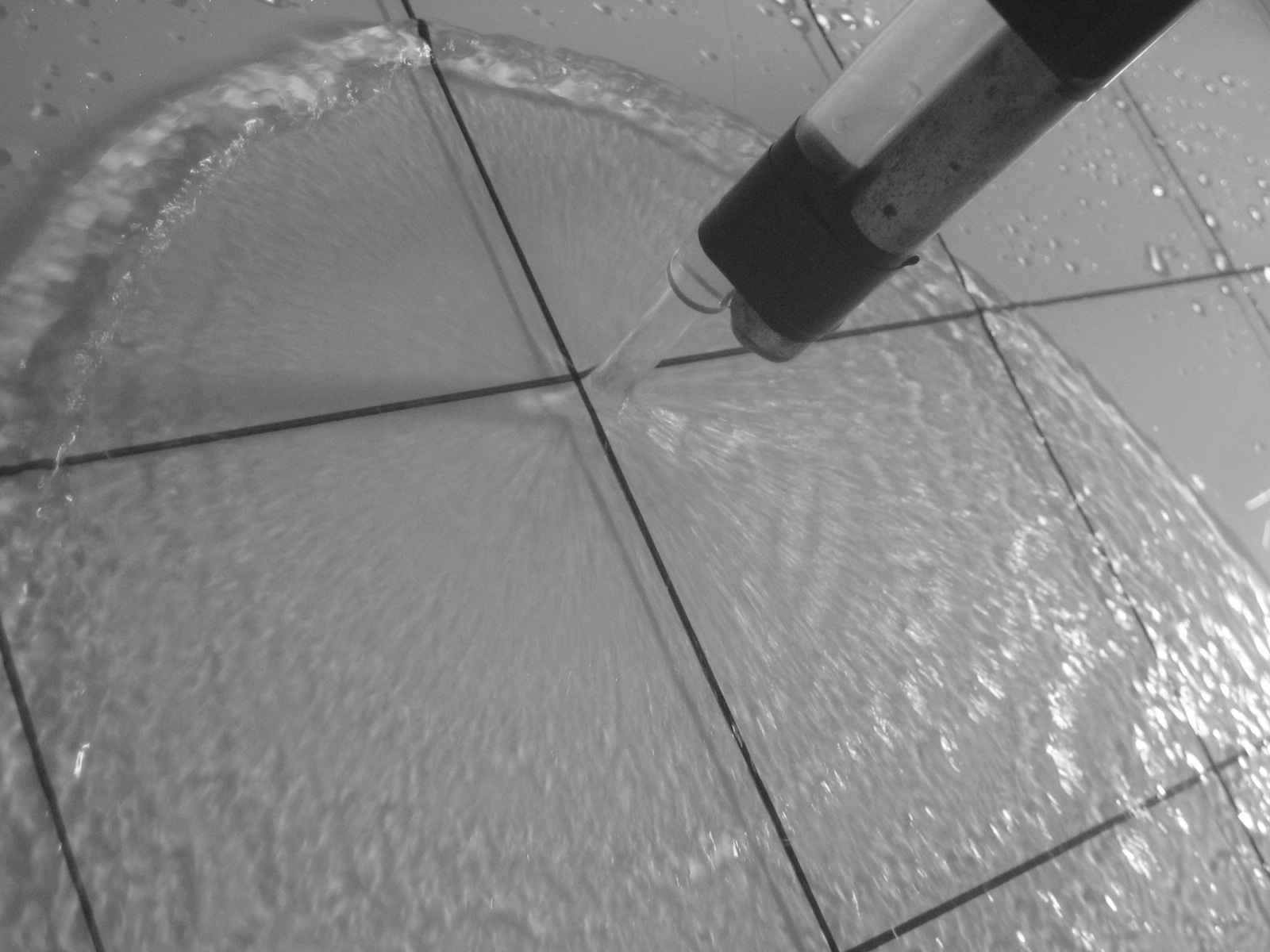}}
  \caption{Impact of a jet on plane inclined at $45\degree$.  The flow rate
    is~$\Q\simeq119\cmcubepersec$ and the rise distance is $\simeq7\cm$.}
  \label{fig:exp}
\end{figure}

Another important case involves the impact of a jet on a plane
inclined with respect to the horizontal, so that there is now a
gravitational force tangential to the plane.  A profile resembling a
parabola is then observed, with a maximum `rise distance' along the
plane (figure~\ref{fig:exp}).  This rise distance can be associated
with the hydraulic jump in the purely horizontal case, although there
are differences.
A full description of this problem is challenging, partly because
of the breaking of axial symmetry, but also due to the necessarily
complex flow in the thick `rim' bounding the inner, shallow region.
For that reason, most existing models are based on inviscid fluid
dynamics (or the opposite extreme of Stokes flow, as
in~\cite{Lister1992}).  \citet{Rienstra1996} first described this
situation in terms of the ballistic motion of fluid particles, leading
to parabolic trajectories with a parabolic envelope or rim.  Such a
ballistic model predicts a rise distance of
order~$\Uc^2/(2\grav\sin\ang)$, where~$\Uc$ is the jet velocity
and~$\grav\sin\ang$ the component of gravity along the plane.  In this
model the particle trajectories are characteristics of a geodesic
equation modified by gravity~\citep{ThiffeaultKamhawi_CCT2007}, and
these characteristics cross.  \citet{Edwards2008} used a `delta-shock'
model to resolve the crossing of characteristics, which lowers the
rise distance to~$5/9$ of the ballistic value whilst maintaining a
near-parabolic outer envelope.  However, in their inviscid model the
rise distance still scales in the same way as for the ballistic
theory.  In their paper \citet{Edwards2008} performed one experiment
(at $90\degree$ inclination angle) and found a significantly lower
value of the rise distance than their theory predicted; they expressed
hope that faster flow rates might capture the inviscid regime, but
this has still to be verified.

In the present paper we account for the discrepancy using a viscous
theory, motivated by observations made in a simple experiment.  The
theory is a straightforward modification of~\citet{Bohr1993}, where
the authors matched inner and outer solutions to viscous shallow-water
equations to predict the radius of a circular hydraulic jump.  To
simplify the treatment, we include a component of the gravitational
force pointing towards the jet while maintaining the assumption of
axial symmetry.  The resulting equation correctly captures the
dependence of the rise distance on both the jet velocity and
inclination angle.  The model, however, is not sufficient to address
other features of the hydraulic jump, such as the shape of the
envelope, or whether the jump closes or has an open (i.e.,
parabolic-like) shape (see~\cite{LebonAPS2008}).  The model does not
agree as well with experiments at slower rates of flow, as
in~\citet{Bohr1993}, or at angles of inclination larger than
about~$60\degree$.

%%%%%%%%%%%%%%%%%%%%%%%%%%%%%%%%%%%%%
\section{Experimental setup and results}
\label{sec:exp}

We performed a simple experimental study on an inclined hydraulic jump
to measure the dependence of the rise distance on flow rate and
inclination angle. A large plexiglas sheet is held clamped over a sink
at a constant angle, so that the water runs directly off at the
edge. This sheet is painted white on one side and marked with a
regular $10\cm$ grid for length calibration. A small pump is used to
supply fresh water at a constant flow rate $\Q$, connected to a
straight glass tube by flexible plastic tubing. The glass tube defines
the nozzle of the jet, which has an inner diameter $d = 0.56\cm$. It
is fastened to a metal rod that can be adjusted so that the jet
strikes the sheet at normal incidence for each value of~$\Q$ and
inclination angle, while the nozzle itself is kept at a $3.5\cm$
distance from the sheet.  The flow rate was measured after each run by
the time taken to fill a one liter beaker for each $\Q$. Based on this
and the nozzle geometry we have an exit velocity $U \simeq 1$--$4$
m/s, with Reynolds number $Re \simeq 6 \times 10^3$--$3 \times 10^4$.

\begin{figure}
  \centerline{\includegraphics[width=.6\textwidth]{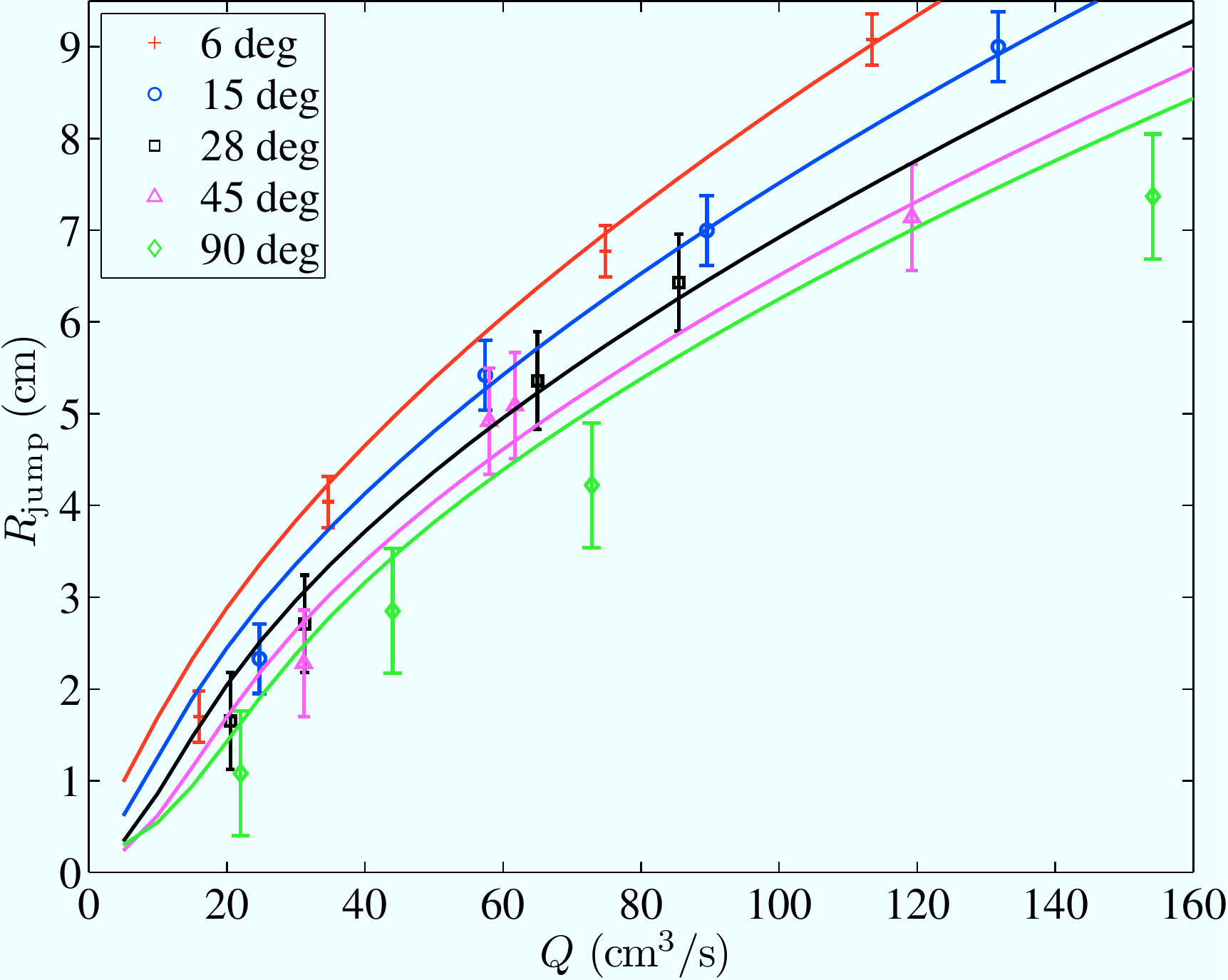}}
  \caption{Maximum rise distance~$\rjump$, measured upslope from the
    center of the jet to the lower edge of the bounding rim.  The
    solid lines come from the theoretical prediction~\eqref{eq:rjump}.
    The theory works well for moderate angles (not near $90\degree$)
    and faster jet speeds.}
  \label{fig:rise}
\end{figure}

We inclined the plexiglas at five different angles $\theta$, measured
with respect to the horizontal: $\theta = 6\degree$, $15\degree$,
$28\degree$, $45\degree$, and $90\degree$.  At each angle, we varied
the flow rate $\Q \simeq 20$--$150\cmcubepersec$.  Finally, from
photographs we measured the rise distance~$\rjump$, which we define as
the distance from the center of the jet to the position of the zenith
of the hydraulic jump (i.e., the lower part of the rim or envelope at
its point of highest rise).  Note that at higher $\Q$ and $\theta$ the
flow becomes more unsteady, and the rise distance has larger error
bars (in those cases we average the rise distance over time).
Figure~\ref{fig:rise} summarizes the results for the
measured~$\rjump$, which increases with velocity and decreases with
$\theta$, as one would expect.  We now present a simple theory to
explain this dependence.

\section{A simple model}
\label{sec:rise}

To capture the maximum rise distance of the fluid along the sloping
surface, we will model the flow as axisymmetric with a radial force,
where the radial force arises from gravity along the slope angle.  The
radial model is suggested by the radial surface waves visible in
figure~\ref{fig:exp}, which appear roughly circular despite the
inclination.  This simplified model will prove sufficient to capture
the scaling of~$\rjump$ with jet velocity and inclination angle.  To
satisfy mass conservation, one can imagine truncating the surface just
past the hydraulic jump, so that fluid can spill out.  This model
resembles a truncated inverted cone, with the jet hitting the apex,
though without the geometrical factors associated with a cone.

Following~\citet{Bohr1993}, we start with the steady, incompressible
Navier--Stokes and mass conservation equations in axisymmetric
cylindrical coordinates, in the boundary layer approximation:
\begin{subequations}
\begin{gather}
  \uc\frac{\pd\uc}{\pd\rc} + \wc\frac{\pd\uc}{\pd\zc} =
  -\grav\cos\ang\frac{d\h}{d\rc} - \grav\sin\ang
  + \visc\frac{d^2\uc}{d\zc^2},\label{eq:NSa}\\
  \frac{\pd\uc}{\pd\rc} + \frac{\uc}{\rc} + \frac{\pd\wc}{\pd\zc} = 0,
  \label{eq:NSb}
\end{gather}
\label{eq:NS}%
\end{subequations}
where~$\uc(\rc,\zc)$ and~$\wc(\rc,\zc)$ are respectively the velocity
components tangent and perpendicular to the solid surface, $\grav$ is
gravitational acceleration, $\h$ is the fluid thickness, and $\visc$
is the kinematic viscosity.  The coordinate~$\rc$ is tangent to the
solid surface, and~$\zc$ is perpendicular to it.  The boundary
conditions at the bottom and top of the fluid are
\begin{subequations}
\begin{align}
  \uc = 0,
  \quad
  \wc = 0,
  \qquad &\text{at } \zc=0;\\
  \frac{\pd\uc}{\pd\zc} = 0,
  \quad
  \wc = \uc\, \frac{d\h}{d\rc},
  \qquad &\text{at } \zc=\h(\rc).
\end{align}
\end{subequations}
Integrating equation~\eqref{eq:NSb} gives the mass conservation equation
\begin{equation}
  \rc\int_0^\h\uc(\rc,\zc)d\zc = \q
  \label{eq:masscons}
\end{equation}
where~$\q=\Q/2\pi$, with~$\Q$ the flow rate of the jet.  In addition,
we must specify the velocity~$\uc(\rc_0,\zc)=\uc_0(\zc)$ and height
$\h(\rc_0)=\h_0$ at a radius~$\rc_0$ larger than the jet radius, since
the boundary layer equations are not valid directly under the jet.
When~$\ang=0$, equations~\eqref{eq:NS} reduce to those
of~\citet{Bohr1993}.

We use the standard hydraulic jump scalings to define dimensionless
`tilde' variables:
\begin{subequations}
\begin{alignat}{3}
  \uc &= \alpha\tilde\uc,\qquad &
  \alpha &= (\ci^{-1/2}\cii^{1/8})&\,&\q^{1/8}\visc^{1/8}\gravt^{3/8},\\
  \wc &= \beta\tilde\wc,\qquad &
  \beta &= &&\q^{-1/4}\visc^{3/4}\gravt^{1/4},\\
  \rc &= \Gamma\tilde\rc,\qquad &
  \Gamma &= (\ci^{1/2}\cii^{-3/8})&\,&\q^{5/8}\visc^{-3/8}\gravt^{-1/8},\\
  \zc &= \delta\tilde\zc,\qquad &
  \delta &= (\cii^{1/4})&\,&\q^{1/4}\visc^{1/4}\gravt^{-1/4},
\end{alignat}
\label{eq:scalings}%
\end{subequations}
except that we included the~$\cos\ang$ dependence in the scalings.
(We will discuss the dimensionless constants~$\ci$ and~$\cii$ below.)
We immediately drop the tildes.

In dimensionless variables, the mass conservation
equation~\eqref{eq:masscons} becomes 
\begin{equation}
  \rc\int_0^\h\uc(\rc,\zc)d\zc = 1.
  \label{eq:masscons2}
\end{equation}
Upon averaging in the~$\zc$ direction, equation~\eqref{eq:NSa} becomes
after integration by parts
\begin{equation}
  2\overline{\uc\frac{\pd\uc}{\pd\rc}}
  + \frac{1}{\rc}\,\overline{\uc^2}
  + \frac{\h'}{\h}\,{\l.\uc^2\r\rvert}_{\zc=\h}
  =
  - \ci \l(\G + \frac{d\h}{d\rc}\r)
  - \frac{\ci}{\cii}\,\frac{1}{\h}\,{\l.\frac{\pd\uc}{\pd\zc}\r\rvert}_{\zc=0}
  \label{eq:main}
\end{equation}
where the $\zc$-average of a function~$F(\rc,\zc)$
is~$\overline{F}(\rc) = (1/\h)\int_0^\h F(\rc,\zc)d\zc$, and we
defined
\begin{equation}
  \G = (\ci^{1/2}\cii^{-5/8})\,\q^{3/8}\visc^{-5/8}\gravt^{1/8}\,\tan\ang\,.
\end{equation}

We now assume the separable form
\begin{equation}
  \uc(\rc,\zc) = \ua(\rc)\ff'(\zc/\h(\rc)),
\end{equation}
where~$\ua=\overline{\uc}$ is the averaged profile, $f$ is a given
function that describes the vertical structure of the thin layer,
with~$\ff(0)=\ff'(0)=\ff''(1)=0$, $\ff(1)=1$.  With this form
for~$\uc$, the mass conservation integral~\eqref{eq:masscons2} becomes
simply~$\ua\h\rr = 1$, which gives a relationship between~$\ua$
and~$\h$.  This allows us to derive the two relations
\begin{equation}
  2\overline{\uc\frac{\pd\uc}{\pd\rc}}
  + \frac{1}{\rc}\,\overline{\uc^2}
  + \frac{\h'}{\h}\,{\l.\uc^2\r\rvert}_{\zc=\h}
  = \ci\,\ua\frac{\pd\ua}{\pd\rc},\qquad\qquad
  {\l.\frac{\pd\uc}{\pd\zc}\r\rvert}_{\zc=0} = \cii\,\frac{\ua}{\h},
  \label{eq:profilerelations}
\end{equation}
with
\begin{equation}
  \ci = \int_0^1\ff'{}^2(\eta)d\eta \qquad\text{and}\qquad
  \cii = \ff''(0).
  \label{eq:cdef}
\end{equation}
The two relations~\eqref{eq:profilerelations} can be used
in~\eqref{eq:main} and in the mass conservation
integral~\eqref{eq:masscons2} to obtain
\begin{equation}
  \ua\ua' + \h' = -\frac{\ua}{\h^2} - \G,\qquad
  \ua\h\rr = 1.
  \label{eq:vh}
\end{equation}
For~$\G=0$ the equations~\eqref{eq:vh} reduce to those of
\citet{Bohr1993}, which are essentially as derived
by~\citet{Kurihara1946} and~\citet{Tani1949b}.  We
combine~\eqref{eq:vh} into one ODE for~$\ua(\rr)$:
\begin{equation}
  \rr(1 - \rr\ua^3)\ua' = -(1 - \G\rr^2\ua - \rr^4\ua^4)\ua\,,
  \label{eq:vode}
\end{equation}
which must be solved together with the flux boundary
condtion~$\ua(\rr_0)=\ua_0$ at the jet radius~$\rr_0$.  Note that
there is a singularity at~$\rc=0$, and another at~$\rc\ua^3=1$.  The
former is not relevant, since we have~$\rc>0$.  The latter will
determine the location of the hydraulic jump, which we will associate
here with the rise distance.  Since~$\G = \G(\Q,\theta)$, this ODE
will have to be solved at each inclination angle and flow rate.

When doing numerical calculations, we will use the parabolic profile
\begin{equation}
  \ff(\eta) = \tfrac32 \eta^2 - \eta^3,
\end{equation}
from which equation~\eqref{eq:cdef} gives~$\ci=6/5$, $\cii=3$.  A more
general approach, for instance using a variable cubic profile as
in~\cite{Bohr1997}, doesn't significantly change the scaling.

Let us examine solutions of~\eqref{eq:vode} for typical experimental
parameters.  For the case shown in figure~\ref{fig:exp},
$\Q\simeq119\cmcubepersec$ and~$\ang=45\degree$.  The viscosity of
water is~$\nu\simeq.01\cmsqpersec$ and~$\grav\simeq980\cmpersecsq$.
The jet radius is $0.28\cm$, and from the flow rate this gives a
velocity~$484\cmpersec$.  Inserting all this into~\eqref{eq:scalings},
we find a horizontal length scale~$\Gamma=11.3\cm$, velocity
scale~$\alpha=9.9\cmpersec$, and~$\G=67$.  In dimensionless form, we
must now integrate~\eqref{eq:vode} from~$\rc_0=(.28/11.3)=.025$
with~$\ua_0=(2/3)(484/9.9)=32.6$. (The $2/3=1/\ff'(1)$ arises from the
choice of a parabolic vertical profile for~$\ff$.)

\begin{figure}
  \centerline{\includegraphics[width=.6\textwidth]{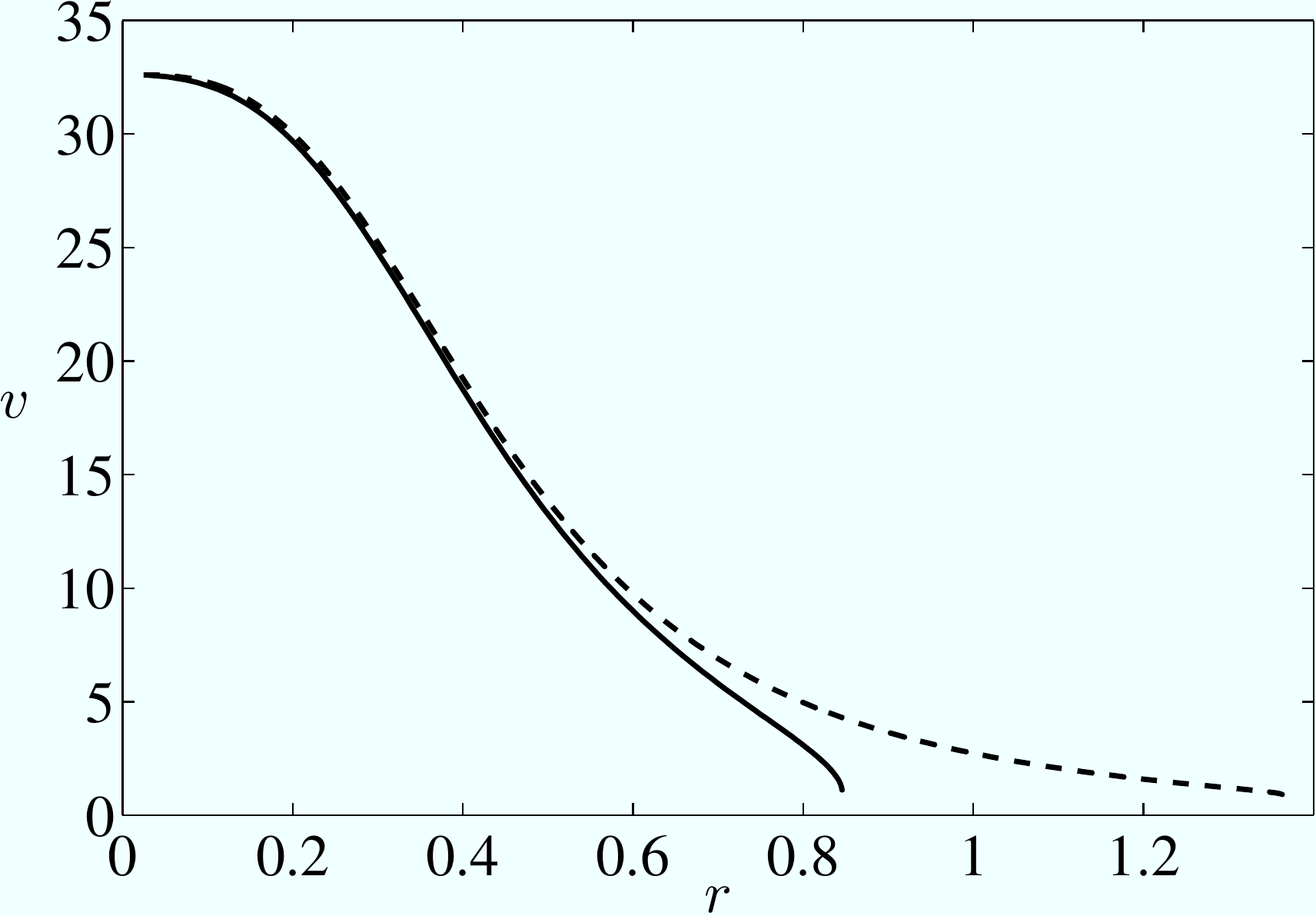}}
  \caption{Solution of equation~\eqref{eq:vode}, with boundary
    condition~$\ua(.025)=32.6$ and parameter~$\G=67$ (---$\!$---),
    corresponding to the values for figure~\ref{fig:exp} in
    dimensionless form.  For comparison, the dashed line (- - - -)
    has~$\G=0$.  Both solutions become singular
    ($\ua'\rightarrow-\infty$) at~$\rcsing=.85$ and $1.36$,
    respectively.}
  \label{fig:vodesol}
\end{figure}

Figure~\ref{fig:vodesol} shows
the numerical solution (solid line), which becomes singular
at~$\rc=\rcsing\simeq.85$.  In dimensional form, the singularity
is~$\rsing\simeq9.6\cm$ from the center of the jet.  The
experimentally-measured value for the lower edge of the hydraulic jump
is~$\rjump\simeq7.2\cm$, with is~$3/4$ of the singularity position.
Note that these values are fairly insensitive to the exact jet
radius~$\rc_0$.  For comparison, the numerical solution for~$\G=0$ is
also indicated as a dashed line in the figure, with a singularity at a
larger value of~$\rc$.

The singularity thus appears to occur somewhat beyond the actual
position of the jump.  To find the position of the jump, we would have
to match to the `outer' solution (small $\ua$) of~\eqref{eq:vode}, and
impose continuity of mass and momentum across the jump, as done
in~\citet{Bohr1993}.  In addition, \citet{Bohr1993} were left with an
extra parameter --- the location of a singularity of the outer
solution --- which they fixed by assuming the jet was striking a plate
of finite extent, and then moving the singularity to the edge of the
plate.  We cannot use this approach here: our hydraulic jump actually
terminates, which by mass conservation means there must be either
backflow or non-axisymmetric flow (or both).  Both these effects
require more powerful theories or the solution of more complex
equations.

On the other hand, since we are only after the scaling of the jump
position, the theory we have is enough to uncover this scaling.
\citet{Bohr1993} observed that their jump location typically occurred
at unit radius (in dimensionless variables), somewhat independently of
what was happening downstream in the outer solution.  This suggests
the following approach: fix the ratio of the jump distance to the
singularity distance from the center of the jet.  In our example
above, that ratio was~$3/4$, but we find $.76$ fits the set of data
slightly better.  Hence, we have in dimensional form~$\rjump \simeq
.76\,\rcsing\,\Gamma$, or after using in~$\Gamma$ parabolic profile
values for the numerical constant,~$\ci^{1/2}\cii^{-3/8}\simeq.73$:
\begin{equation}
  \rjump \simeq .55\,\,
  \rcsing(\G,\rc_0,\ua_0) \times \q^{5/8}\visc^{-3/8}\gravt^{-1/8}.
  \label{eq:rjump}
\end{equation}
Here~$\rcsing(\G,\rc_0,\ua_0)$ is obtained by solving the
dimensionless ODE~\eqref{eq:vode} with initial condition at the jet
radius~$\ua(\rc_0)=\ua_0$, and~$\ua_0$ is obtained from the
dimensionless flow rate by~$\ua_0=(2/3)(\Q/\pi\rc_0^2)$, where
the~$2/3$ is for a parabolic vertical profile.

In figure~\ref{fig:rise} we compare formula~\eqref{eq:rjump} with
experiments at various inclination angles, with the jet always
striking the plane normally.  We emphasize that the numerical
prefactor in~\eqref{eq:rjump} is fixed, so we are not fitting each
data set individually.  The theory agrees well with experiments in
both velocity and angle, except at low flow rates and at $90\degree$
angle.  The low flow rate disagreement is not troubling, since it
falls outside the theory as pointed out by \citet{Bohr1993}.  The
$90\degree$ theoretical curve stands out, since it is very close to
the theoretical curve for $45\degree$.  However, the trend of the
$90\degree$ curve with~$\Q$ is still captured.

\begin{figure}
  \centerline{\includegraphics[width=.6\textwidth]{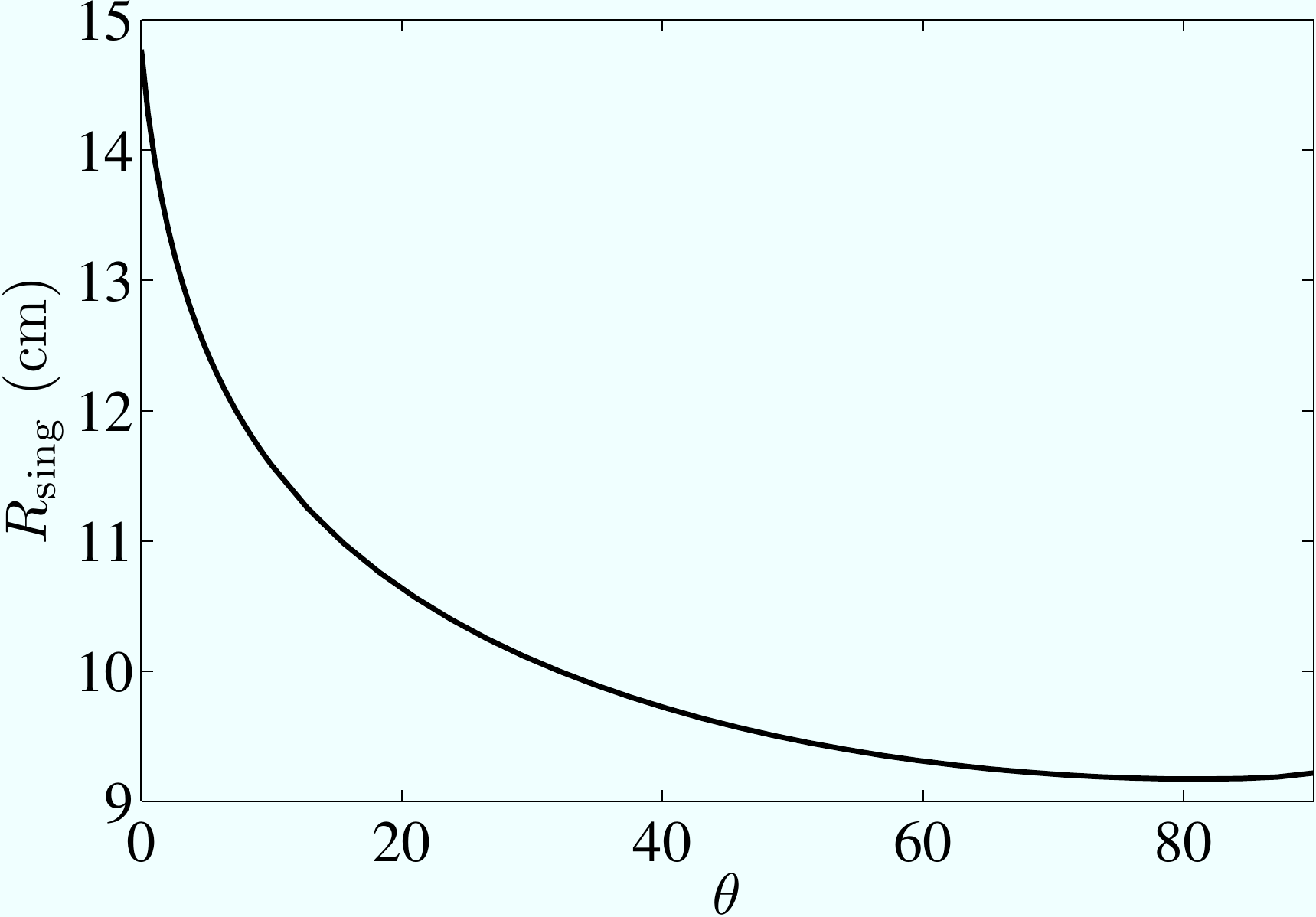}}
  \caption{Position of the singularity in equation~\eqref{eq:vode} as
    a function of inclination angle, for a fixed flow
    rate~$\Q=119\cmpersec$.  The curve has a minimum
    around~$80\degree$.}
  \label{fig:vsing}
\end{figure}

Figure~\ref{fig:vsing} shows the singularity position as a function of
angle, for a fixed flow rate $119\cmpersec$.
The curve actually has a minimum around~$80\degree$, which is
unphysical: physically we expect the rise distance to decrease
monotonically with angle, though at these large angles of inclination
the flow is too unsteady for accurate measurement of the rise
distance.  We conclude that the theory breaks down for steeper angles:
probably backflow becomes important, and the assumption that the the
outer solution only weakly affects the jump position breaks down.
According to figure~\ref{fig:vsing}, the rise distance ceases to
increase significantly after about~$60\degree$; however
figure~\ref{fig:rise} shows a significantly higher rise distance
for~$90\degree$ inclination.  This suggests that the present theory
works well for angles~$\ang\lesssim60\degree$.

\section{Conclusions}

We have presented simple experiments of the impact of a jet on an
inclined plane, and noted the dependence of the maximum rise distance
on both flow rate and angle.  Though a complete description of this
problem is daunting, our simple radial model captures the dependence
remarkably well, though less so at smaller flow rates and larger
angles.  A complete theory would, of course, describe the thickness of
the bounding rim, but also predict the critical angle at which the
hydraulic jump changes from closed to open \citep{LebonAPS2008}.  Our
simple model verifies the need to include viscosity to capture the
rise height at these modest flow rates, as pointed out
in~\citet{Edwards2008}.

\begin{acknowledgments}
  We thank Claudia Cenedese for graciously allowing us to use her lab,
  as well as Shreyas Mandre, Cecilia Ortiz-Duenas and J.~B. Keller for
  helpful discussions.  J-LT and AB are grateful for the hospitality
  of the 2008 Summer Program in Geophysical Fluid Dynamics (supported
  by NSF and ONR) at WHOI, where this work began, and the Institute
  for Mathematics and its Applications (supported by NSF).  J-LT was
  supported by NSF under grant DMS-0806821.
\end{acknowledgments}

%\bibliographystyle{jfm}
%\bibliography{bib/journals_abbrev,bib/articles,lebon}

\end{document}